# Quantum random number generator based on room-temperature single-photon emitter in gallium nitride

Qing Luo,[1, 2] Zedi Cheng,[1] Junkai Fan,[1] Lijuan Tan,[1, 2] Haizhi Song,[1, 2] Guangwei Deng,[1, ‡] You Wang,[1, 2, †], and Qiang Zhou,[1, 3, *]

[1] *Institute of Fundamental and Frontier Sciences, University of Electronic Science and Technology of China, Chengdu 610054, China*
[2] *Southwest Institute of Technical Physics, Chengdu 610041, China*
[3] *School of Optoelectronic Science and Engineering, University of Electronic Science and Technology of China, Chengdu 610054, China*
[‡] *gwdeng@uestc.edu.cn;* [†] *youwang_2007@aliyun.com;* [*] *zhouqiang@uestc.edu.cn*

We experimentally demonstrate a real-time quantum random number generator by using a room-temperature single-photon emitter from the defect in a commercial gallium nitride wafer. Thanks to the brightness of our single photon emitter, the raw bit generation rate is ~1.8 MHz, and the unbiased bit generation rate is ~420 kHz after von Neumann's randomness extraction procedure. Our results show that commercial gallium nitride wafer has great potential for the development of integrated high-speed quantum random number generator devices.

## I. INTRODUCTION

Random numbers have been widely applied in science and engineering [1, 2]. Generally, there are two methods of generating random numbers: pseudorandom number generators (PRNGs) and true number generators (TRNGs) [1]. For PRNGs, algorithmic methods are used to produce random-looking numbers, which have been applied in computer simulations [3-5]. For TRNGs, physical methods are used to produce true random numbers, which play a key role in quantum cryptography and fundamental science experiments [6, 7]. By exploiting the inherent randomness of a quantum mechanics system, quantum random number generators (QRNGs) can generate true and unpredictable random numbers [1]. QRNGs have been demonstrated in various ways, including single-photon detection [2, 8, 9], vacuum fluctuations [10-13], laser phase noise [14, 15], amplified spontaneous emission (ASE) [16-18], and quantum non-locality [19-22], etc.

One typical QRNG based on the single-photon detection is the branching path generator, where the randomness entropy is obtained by detecting the path superposition of a single photon [1]. A single photon passes through a balanced beam splitter with equal transmittance and reflectance, and the single-photon quanta cannot be divided into two parts. One can define a state $|1\rangle_t|0\rangle_r$ to represent a photon in the transmission path, and a state $|0\rangle_t|1\rangle_r$ with the photon in the reflection path. Thus, a path superposition state can be prepared as expressed by

$$|\phi\rangle = \frac{|1\rangle_t|0\rangle_r + |0\rangle_t|1\rangle_r}{\sqrt{2}}. \quad (1)$$

With two single photon detectors (SPD1 and SPD2) placed at two outputs of the beam splitter, a binary random bit of '0' or '1' is obtained when a photon is detected by SPD1 or SPD2 respectively.

In the branching path generator, both the weak coherent light and the genuine single photon source can be utilized as the randomness entropy source [23]. Due to the Poisson photon statistic of the weak coherent state, the mean photon number of the coherent state must be attenuated to around 0.1 to be thought as a single photon level [24]. On the other hand, the mean photon number from a genuine single photon source can achieve 1. Moreover, a QRNG based on the genuine single photon source exhibits a higher entropy per raw bit than the weak coherent state one, which has been theoretically analyzed by Oberreiter et al. [25].

In this paper, we experimentally demonstrate a real-time branching path generator, which is based on the bright room-temperature single-photon emission from a single defect in gallium nitride (GaN) [26]. In our experiment, a single photon emission rate of 2 MHz - with the second-order auto-correlation of $0.36 \pm 0.01$ - has been achieved, which results in a raw-bits generation rate of 1.8 MHz. Furthermore, we adopt the von Neumann's procedure to extract the unbiased random bits in real time, and obtain the unbiased bit generation rate of 420 kHz. Our results pave the way for the development of GaN-chip based QRNG devices.

## II. EXPERIMENTS AND RESULTS

### 1. Performance of single photon emitter (SPE)

The single photon emitter from the defect center in a GaN layer grown on a patterned sapphire substrate (PSS) is used for randomness generation. The defect center is optically excited and located by a homemade confocal microscopy [27]. The experimental setup is shown in Fig. 1a. A continuous wave 532 nm laser is used to excite the single emitter. To clean the mode of the laser, the 532 nm laser beam is coupled into an optical single-mode fiber, and then injected into the microscopy system. The laser beam is reflected by a dichroic mirror with the cut-off wavelength of 560 nm, and then focused onto the GaN wafer by a $100 \times$ objective lens with the NA of 0.9. The emitted single-photon fluorescence is also collected by the same objective lens in the confocal configuration, and then pass through the dichroic mirror towards the detection system. To suppress the unwanted scattered excitation light and stray light, the fluorescence is filtered by a long-pass filter with a cutoff wavelength of 600 nm, and then is tightly focused into a single-mode fiber by a lens with the focal length of 15 mm. The confocal fluorescence map scanning is realized by using a nanopositioner with the position accuracy of ~20 nm.

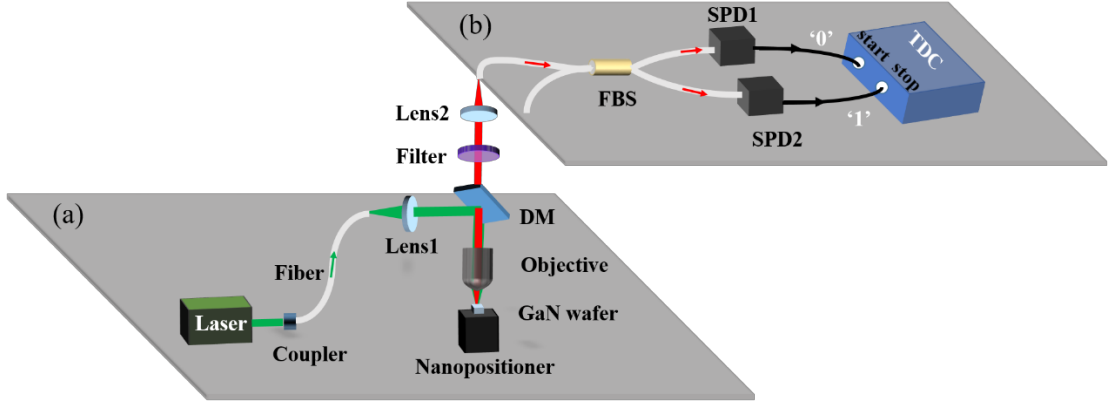

FIG. 1. Schematic illustration of branching path random bit generator. (a) Experimental setup of the homemade confocal scanning microscopy, which is used to excite the single photon emitter in the GaN wafer, and to collect the single-photon fluorescence. DM: dichroic mirror. (b) HBT configuration, used to exam the single photon purity of emitted photons and to generate for the binary random number generation. FBS: fiber beam splitter, SPD: single photon detector, TDC: time-to-digital converter.

To obtain the single photon purity of photons from the selected emitter, we measure the value of second-order auto-correlation function - $g^{(2)}(\tau)$ - of the photons through a Hanbury Brown and Twiss (HBT) setups [28] as shown in Fig. 1b, where $\tau$ is the time delay- the sum of optical time delay and electrical time delay - between the two output paths of the HBT setups. The collected photons are equally divided by a 50:50 fiber beam splitter (FBS). The photons from two output ports of the FBS are respectively detected by two SPDs. Detection signals from the two SPDs are fed into a time-to-digital converter (TDC), in which the coincidence counts of the two SPDs are recorded. Then, the $g^{(2)}(\tau)$ is obtained based on the coincidence counts and the single-photon detection from the two SPDs.

In our experiment, we study the saturation behavior of the emitter to obtain the optimal excitation power for the QRNG source firstly. As shown in figure 2(a), the photon emission rate increases with the excitation power. The saturation curve can be obtained by fitting with the mathematical model as given by [26, 29]

$$I(P) = \frac{I_\infty P}{P+P_s} ,  \quad (2)$$

where $I(P)$ is the measured intensity count rate, $P$ is the excitation power, $P_s$ is the saturation power, and $I_\infty$ is the maximum emission rate, respectively. Based on the fitting of our experimental data, the saturation power $P_s = 3$ mW has been obtained in our experiment. Under the laser excitation with the saturation power of 3 mW - the star dot in Fig. 2a, the photon count rate of about 2 MHz is reached at room temperature (RT). With the background fluorescence of about 600 kHz, the signal-to-noise ratio (SNR) of the emitter is measured as $\rho = 0.75$. The $g^{(2)}_{raw}(\tau)$ of the emitter has been recorded. And the corrected $g^{(2)}(\tau)$ can be obtained by using a background correction equation from Ref. [30],

$$g^2(\tau) = \frac{[g^2_{raw}(\tau)-(1-\rho^2)]}{\rho^2} , \quad (3)$$

where ρ is the SNR of SPE. Figure 2b shows the corrected $g^{(2)}(\tau)$ of the SPE under CW laser excitation at RT. The obtained value of $g^{(2)}(0)$ is $0.36 \pm 0.01$, which is below the classical threshold of 0.5 - the dashed line in Fig. 2b. Therefore, we can consider our emitter as a SPE.

**2. Performance of quantum random number generation**

The randomness entropy source comes from the measurement of the path superposition state. The experimental setup for the binary random number generation is also shown in Fig. 1b. A stream of single photons is guided into a FBS with the coupling ratio of 50:50 and the operation wavelength range of $670 \pm 75$ nm. The second input arm of the FBS is blocked, which means that a vacuum state is inputted. The probability whether a photon is detected in either output path is related to the vacuum state. The transmitted and reflected photons are detected by the SPD1 and SPD2, respectively. The dead time of each SPD is ~30 ns, and the dark count is ~50 Hz. Electrical pulses of two SPDs are then inputted to two channels of the TDC. The time bin of the TDC is set to 176 ps. Then, the time-stamps of these electrical pulses, i.e., all detection events, can be recorded by such a TDC. Clicks in SPD1 are recorded as bits of '0', and clicks in SPD2 are recorded as bits of '1'. A sequence of raw random bits can be acquired by use of the experimental setup as shown in Fig. 1. For post-processing, the von Neumann's de-biasing procedure is employed to extract the unbiased random numbers in our study [23, 31].

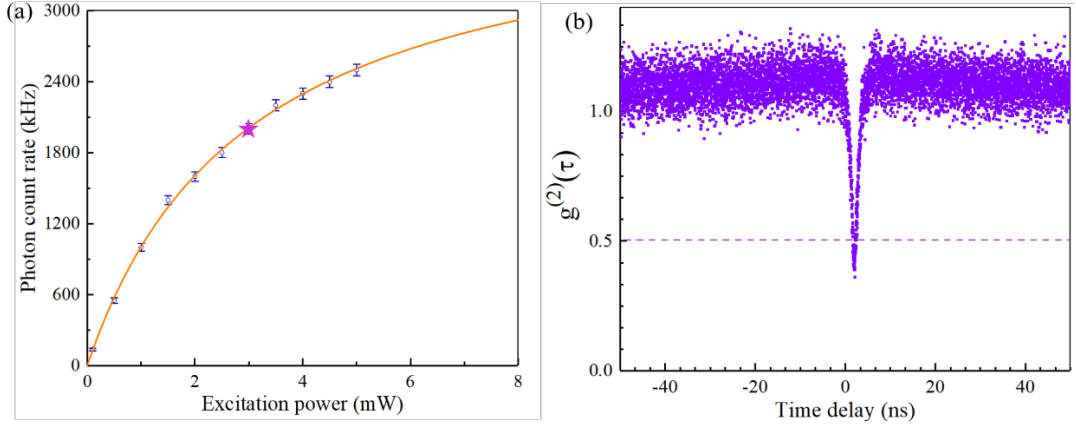

FIG. 2. Properties of our SPE. (a) Saturation behavior of the SPE. The optimal excitation power for the emitter is 3 mW, as denoted by the saturation curve (red). (b) Second-order auto-correlation measurement of the defect center excited by a 532 nm CW laser with the power of 3 mW - the star dot in (a). The dashed line is the classical threshold of 0.5.

For every pair of generated bits, the results of '00' and '11' are discarded, while the results of '01' and '10' are replaced by '0' and '1', respectively. By using the von Neumann's de-biasing procedure, some dependence of two adjacent bits can be eliminated. Moreover, as the von Neumann's de-biasing algorithm is used directly and simply in our data acquisition program, the unbiased random bits can be obtained in real time.

In our proof of principle demonstration, we approximately acquire 6 G raw bits and 1 G unbiased bits of both '0' and '1' in the course of 3784 seconds. The generation rates of the raw bits and unbiased bits are 1.8 MHz and 420 kHz, respectively. For the raw bits, the probabilities of '0' and '1' bit are $P(0) = 0.5814$, $P(1) = 0.4186$, respectively. For the unbiased bits, the probabilities of '0' and '1' bit are $P(0) = 0.5007$, $P(1) = 0.4993$, respectively. We calculate the autocorrelation coefficients of the raw and unbiased bits. As shown in Fig. 3, the autocorrelation coefficients of the random sequence are reduced after the von Neumann's de-biasing procedure. Then, 6000 1-Mbits raw sequences and 1000 1-Mbits unbiased sequences are respectively tested by NIST Statistical Test Suite which is commonly known [32, 33]. There are 15 sub-tests for the analyses of the statistical properties of random numbers. The passing proportions of the each NIST sub-test can be determined within a range as given by $p \pm 3\sqrt{p(1-p)/N}$, where $p = 1 - \partial$, the significance level $\partial$ is set to 0.01 as suggested by the test suite, and N is the number of test trials, respectively. From the test results shown in Fig. 4, while the raw bits fail several tests, the unbiased bits pass all sub-tests with a proportion value greater than 0.99 and less than 1, and the results are considered to be acceptable because a series of *p*-values obtained in the sub-tests are above the suggested significance level of 0.01.

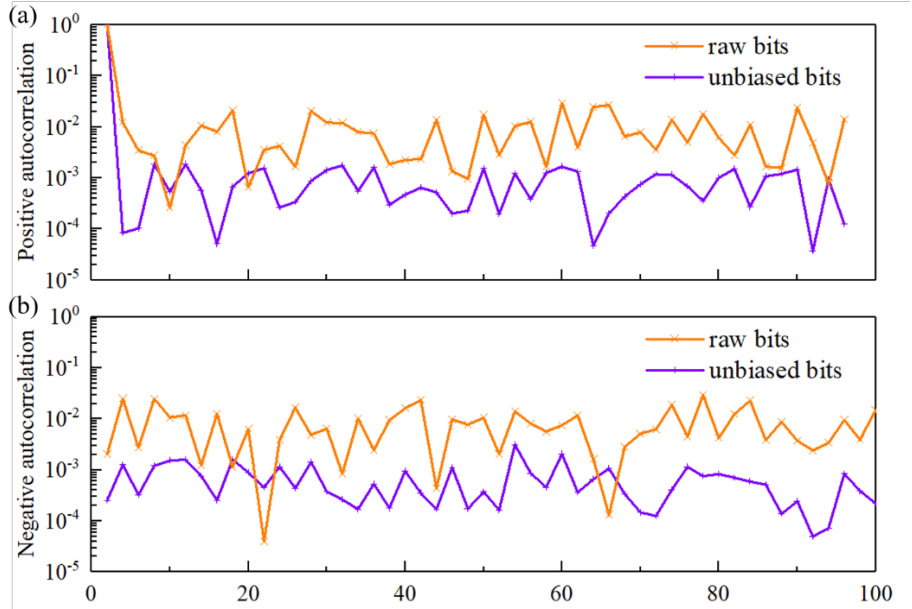

FIG. 3. Positive (a) and negative (b) autocorrelation coefficient curves within 100-bits delay for the raw and unbiased bits. The autocorrelation coefficients decrease after post-processing.

each other. Another factor is the detection responses of two detectors. Since the PL count rate of the SPE is ~2MHz, i.e., the average waiting time between two photon detection events is about 500 ns, which is much longer than the dead times (~30 ns) of two SPDs. Thus, in most cases, every two adjacent bits can be independent from each other. However, for conservative consideration of the randomness, two adjacent bits from different detectors might be dependent sometimes. This happens when two subsequent photons are detected within the interval shorter than the dead time of detectors. One can deduce that these two photons must be detected by two different detectors. On this condition, the results of '01' or '10' are not independent. Hence, the von Neumann's randomness extraction procedure is used to ensure the true amount of randomness at a certain level of accuracy.

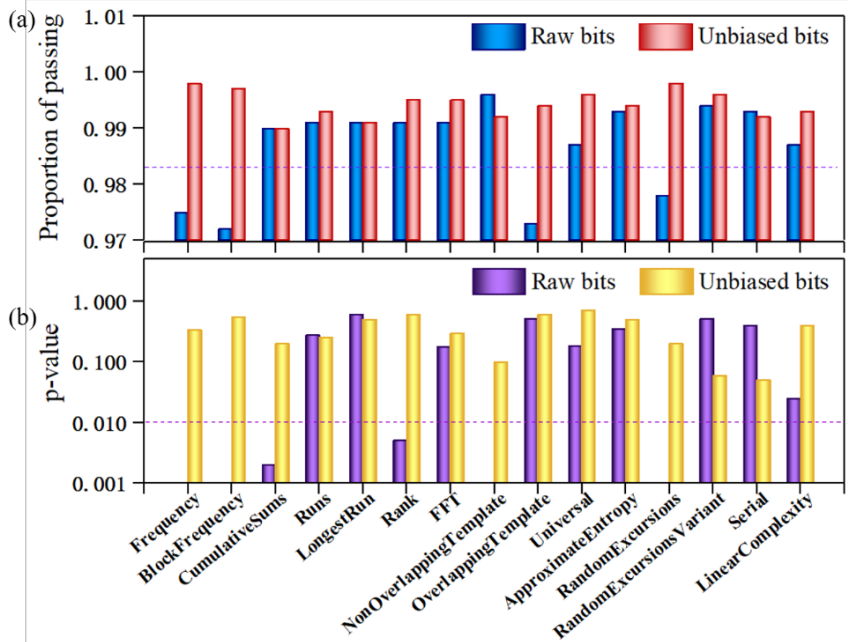

FIG. 4. NIST test results for the raw and unbiased bits. The unbiased bits pass all the tests with proportion values ≥ 0.99 and *p*-values ≥ 0.01.

The raw random bits are biased to some extent, which means that the bits of '0' and '1' do not strictly obey a uniform distribution. One of the main influence factors is the unbalance of the beam-splitting ratio. This could be due to the wavelength range of single-photon fluorescence is wider than the operation range of the FBS. It can be improved either by narrowing the fluorescence or by developing FBS with wider operation range. Fortunately, the biased beam splitter does not necessarily introduce any memory in the system [34], so the path choices can still be independent from

### III. DISCUSSION AND CONCLUSION

For genuine single-photon based QRNGs, the photon count rate of the SPE is a limiting factor for the bit generation rate. We note[1] that QRNGs based on single photons from the solid-state defects in diamond and hexagonal boron nitride have been demonstrated by Chen X et al. [34] and White et al. [35], respectively. In comparison with these kinds of QRNGs, our QRNG based on a bright SPE from the defect center in GaN exhibits the advantage of a faster bit generation rate. The brightness of our SPE is up to about 2MHz, which determines the relatively high bit generation rate of about 420 kHz.

Here we focus on discussing the high brightness of our SPE. Primarily, the PSS of the GaN LED wafer could lead to the 2-fold intensity enhancement as compared with the flat substrate [29]. Besides, our SPE works under a relatively high excitation power - the saturation excitation power of 3 mW. Generally, the emission rate of the SPE increases with the excitation power, whereas the shape of anti-bunching curves becomes narrower when the power increases, which indicates that the quantum randomness entropy would decrease [34]. Thus, for a trade-off between the bit generation rate and the quantum randomness, our SPE is excited under the saturation excitation power. However, the PSS structure and high excitation power lead to not only the high brightness of the SPE but also a lot of background noise from the stray light. The background noise can interfere with both the measurement of $g^{(2)}(\tau)$ and the generation of random bits. For the verification of our SPE, the pure $g^{(2)}(\tau)$ of 0.36 is extracted from the background noise. For the output of our QRNG, the background noise might introduce the dependence of two adjacent bits [34], while such dependence can be restrained by the von Neumann's extraction procedure.

It should be noted that the photon count rate of our SPE is far smaller than the maximum theoretical emission rate of 100 MHz [25], which implies that the collection efficiency of emitted light to free-space optics is quite low. There are mainly two aspects to further improve the photon count rate of the SPE used in QRNGs. On one hand, enhancement in collection efficiency can be achieved by coupling the defect centers to photonic structures, including solid immersion lenses [36], nanopillars [37], and circular bulls eye gratings [38]. A 15-fold enhancement in the detection rate could be obtained [39]. On the other hand, the emission rate of SPEs can be increased by efficiently coupling defect centers to integrated optical cavities via the Purcell effect [39]. With a GaN based SPE coupled in a waveguide coupled resonator and an integrated single-photon detector patterned on top of GaN waveguides, a fast on-chip QRNG device are expected to be explored.

In conclusion, we have demonstrated a fast QRNG based on a bright SPE from the defect center in GaN. The fast bit generation rate is attributed to the high brightness of the SPE. In our work, the brightness of the SPE is 2 MHz, and the unbiased bit generation rate is up to about 420 kHz with a real-time simple von Neumann's de-biasing procedure. Our results pave the way to a fast on-chip QRNG device based on GaN defect centers.

## ACKNOWLEDGEMENTS

This work is supported by National Key R&D Program of China (2018YFA0307400); National Natural Science Foundation of China (NSFC) (U19A2076, 61775025, 91836102, 61704164). We thank Dr. Junfeng Wang and Mr. Qiang Li for supports on the establishment of our experimental platform.

---

[1] The generator in Ref. [34] is continuously operated over the course of one week, and the generation rate of raw bits is ~100 kHz, which exhibits the relatively low-efficiency and is non-real-time. The total photon count rate of SPE in Ref. [35] is ~70 kHz.